\documentclass[12pt]{article}
\usepackage{paper2e}
\usepackage{epsfig}
\usepackage{latexsym}

\newcommand{\roughly}[1]%
    {{\mathrel{\raise.3ex\hbox{$#1$\kern-.75em\lower1ex\hbox{$\sim$}}}}}
\newcommand{\lsim}{\mathrel{\roughly<}}

\begin{document}
\begin{titlepage}

\preprint{UTAP-564, RESCEU-20/06}

\title{Accelerating Universe and Cosmological Perturbation in the Ghost
 Condensate}

\author{Shinji Mukohyama}

\address{Department of Physics and 
Research Center for the Early Universe,\\
The University of Tokyo, Tokyo 113-0033, Japan}

\begin{abstract}
 In the simplest Higgs phase of gravity called ghost condensation, an
 accelerating universe with a phantom era ($w<-1$) can be realized
 without ghost or any other instabilities. In this paper we show how to
 reconstruct the potential in the Higgs sector Lagrangian from a given
 cosmological history ($H(t)$, $\rho(t)$). This in principle allows us
 to constrain the potential by geometrical information of the universe
 such as supernova distance-redshift relation. We also derive the
 evolution equation for cosmological perturbations in the Higgs phase of
 gravity by employing a systematic low energy expansion. This formalism
 is expected to be useful to test the theory by dynamical information of
 large scale structure in the universe such as cosmic microwave
 background anisotropy, weak gravitational lensing and galaxy
 clustering. 
\end{abstract}

\end{titlepage}

\section{Introduction}
\label{sec:intro}

Acceleration of the cosmic expansion today is one of the greatest
mysteries in both cosmology and fundamental
physics~\cite{Riess:1998cb,Perlmutter:1998np}. Assuming that Einstein's
general relativity is the genuine description of gravity all the way up 
to cosmological distance and time scales, the so called concordance 
cosmological model requires that about $70\%$ of our universe should be
some sort of energy with negative pressure, called dark energy. However,
since the nature of gravity at cosmological scales has never been probed
directly, we do not know whether the general relativity is really
correct at such infrared (IR) scales. Therefore, it seems natural to
consider modification of general relativity in IR as an alternative to
dark energy. Dark energy, IR modification of gravity and their
combination should be tested and distinguished by future observations
and
experiments~\cite{Linder:2005in,Ishak:2005zs,Bertschinger:2006aw,Shirata:2005yr,Yamamoto:2006yv}.

From the theoretical point of view, however, IR modification of general
relativity~\cite{Brans:1961sx,Dicke:1961gz,Bergmann:1968ve,Clayton:1998hv,Clayton:1999zs,Clayton:2001rt,Moffat:2002kj,Drummond:1999ut,Drummond:2001rj,Carroll:2003wy} 
is not an easy subject. Most of the previous proposals are one way or
another scalar-tensor theories of gravity, and are strongly constrained
by e.g. solar system experiments~\cite{Will:1993ns} and the theoretical
requirement that ghosts be
absent~\cite{Cline:2003gs,DeFelice:2006pg,Calcagni:2006ye}. The massive
gravity theory~\cite{Fierz:1939ix} and the Dvali-Gabadadze-Porrati (DGP)
brane model~\cite{Dvali:2000hr} are much more interesting IR
modification of gravity, but they are known to have macroscopic UV 
scales~\cite{Arkani-Hamed:2002sp,Luty:2003vm}. A UV scale of a theory is
the scale at which the theory breaks down and loses its
predictability. For example, the UV scale of the $4D$ general relativity
is the Planck scale, at which quantum gravity effects are believed to 
become important. Since the Planck scale is microscopic, the general
relativity maintains its predictability at essentially all scales we can
directly probe. On the other hand, in the massive gravity theory and the
DGP brane model, the UV scale is macroscopic. For example, if the scale
of IR modification is the Hubble scale today or longer then the UV scale
would be $\sim 1,000km$ or longer. At the UV scale an extra degree of
freedom, which is coupled to matter, becomes strongly coupled and its
quantum effects cannot be ignored. This itself does not immediately
exclude  those theories, but means that we need UV completion in order
to predict what we think we know about gravity within $\sim
1,000km$. Since this issue is originated from the IR modification and
the extra degree of freedom cannot be decoupled from matter, it is not
clear whether the physics in IR is insensitive to unknown properties of
the UV completion. In particular, there is no guarantee that properties
of the IR modification of gravity will persist even qualitatively when
the theories are UV completed in a way that they give correct
predictions about gravity at scales between $\sim 1,000km$ and 
$\sim 0.1mm$.

Ghost condensation is an analogue of the Higgs mechanism in general
relativity and modifies gravity in IR in a way that avoids the
macroscopic UV scale~\cite{Arkani-Hamed:2003uy}~\footnote{
See
e.g.\cite{Rubakov:2004eb,Dubovsky:2004sg,Holdom:2004yx,Bluhm:2004ep,Bluhm:2006im,Kirsch:2005st,Cheng:2006us} 
for other related proposals. 
}. In ghost condensation the theory is expanded around a background
without ghost and the low energy effective theory has a universal
structure determined solely by the symmetry breaking pattern. While the
Higgs mechanism in a gauge theory spontaneously breaks gauge symmetry,
the ghost condensation spontaneously breaks a part of Lorentz
symmetry~\footnote{
Lorentz violation in particle physics has been an active field of
research. See
e.g.~\cite{Colladay:1998fq,Bluhm:2005uj,Mattingly:2005re,Amelino-Camelia:2005qa,Vucetich:2005ra} 
and references therein. Gravity in the presence of a vector field with
fixed norm is also extensively
studied~\cite{Jacobson:2000xp,Eling:2003rd,Jacobson:2004ts,Jacobson:2005bg,Kanno:2006ty}.
For relation between this theory and the gauged ghost condensation, see 
Appendix of ref.~\cite{Cheng:2006us}. 
}
since this is the symmetry relevant to gravity. In a gauge theory the
Higgs mechanism makes it possible to 
give a mass term to the gauge boson and to modify the force law in a
theoretically controllable way. Similarly, the ghost condensation gives
a ``mass term'' to the scalar sector of gravity and modifies
gravitational force in the linearized level even in Minkowski and de
Sitter spacetimes. The Higgs phase of gravity provided by the ghost
condensation is simplest in the sense that the number of Nambu-Goldstone
bosons associated with spontaneous Lorentz breaking is just one and that
only the scalar sector is essentially modified.

In ghost condensation the linearized gravitational potential is modified
at the length scale $r_c$ in the time scale $t_c$, where $r_c$ and $t_c$
are related to the scale of spontaneous Lorentz breaking $M$ as
%
\begin{equation}
 r_c \simeq \frac{M_{\rm Pl}}{M^2}, \quad
 t_c \simeq \frac{M_{\rm Pl}^2}{M^3}. 
\end{equation}
Note that $r_c$ and $t_c$ are much longer than $1/M$. The way gravity is
modified is peculiar. At the time when a gravitational source is turned
on, the potential is exactly the same as that in general
relativity. After that, however, the standard form of the potential is
modulated with oscillation in space and with exponential growth in
time. This is an analogue of Jeans instability, but unlike the usual
Jeans instability, it persists in the linearized level even in Minkowski
background. The length scale $r_c$ and the time scale $t_c$ above are
for the oscillation and the exponential growth, respectively. At the
time $\sim t_c$, the modification part of the linear potential will have
an appreciable peak only at the distance $\sim r_c$. At larger
distances, it will take more time for excitations of the Nambu-Goldstone
boson to propagate from the source and to modify the gravitational
potential. At shorter distances, the modification is smaller than at the
peak position because of the spatial oscillation with the boundary
condition at the origin. The behavior explained here applies to
Minkowski background, but in ref.~\cite{Arkani-Hamed:2003uy} the
modification of gravity in de Sitter spacetime was also analyzed. It was
shown that the growing mode of the linear gravitational potential
disappears when the Hubble expansion rate exceeds a critical value
$H_c\sim 1/t_c$. Thus, the onset of the IR modification starts at the
time when the Hubble expansion rate becomes as low as $H_c$.

If we take the $M/M_{\rm Pl}\to 0$ limit then the Higgs sector is
completely decoupled from the gravity and the matter sectors and, thus,
the general relativity is safely recovered. Therefore, cosmological and
astrophysical considerations in general do not set a lower bound on the
scale $M$ of spontaneous Lorentz breaking, but provide upper bounds on
$M$. If we trusted the linear approximation for all gravitational
sources for all times then the requirement $H_c \lsim H_0$ would give
the bound $M\lsim (M_{\rm Pl}^2H_0)^{1/3}\simeq 10 MeV$, where $H_0$ is
the Hubble parameter today~\cite{Arkani-Hamed:2003uy}. However, for
virtually all interesting gravitational sources the nonlinear dynamics
dominates in time scales shorter than the age of the universe. As a
result the nonlinear dynamics cuts off the Jeans instability of the
linear theory, and allows  $M\lsim 100 GeV$~\cite{Arkani-Hamed:2005gu}.

Many other aspects of ghost condensation have been explored. They
include a new spin-dependent force~\cite{Arkani-Hamed:2004ar},  
a qualitatively different picture of inflationary de Sitter
phase~\cite{Arkani-Hamed:2003uz,Senatore:2004rj}, 
effects of moving sources~\cite{Dubovsky:2004qe,Peloso:2004ut}, 
nonlinear dynamics~\cite{Krotov:2004if,Arkani-Hamed:2005gu}, 
properties of black holes~\cite{Frolov:2004vm,Mukohyama:2005rw,Dubovsky:2006vk},
implications to galaxy rotation curves~\cite{Kiselev:2004bh,Kiselev:2005sv,Kiselev:2006gc},
dark energy models~\cite{Piazza:2004df,Krause:2004bu,deRham:2006pe},
other classical dynamics~\cite{Anisimov:2004sp,Mann:2005jz}, 
attempts towards UV
completion~\cite{Graesser:2005ar,O'Connell:2006de},
and so on.

In the simplest setup of the ghost condensation an exact shift symmetry
is assumed and there is no potential term in the Lagrangian of the Higgs
sector. As a result the Higgs sector behaves like a cosmological
constant plus cold dark matter for homogeneous, isotropic background 
evolution~\cite{Arkani-Hamed:2005gu}. If the shift symmetry is not exact
but is softly broken then a shallow potential is allowed in the Higgs
sector Lagrangian. Recently, Creminelli, et.al~\cite{Creminelli:2006xe}
showed that the ghost condensation with softly broken shift symmetry can
violate the null energy condition without any instabilities~\footnote{
Another model which violates the null energy condition without UV 
instabilities is proposed in \cite{Rubakov:2006pn}. It is interesting to
notice that this model also breaks Lorentz symmetry. 
}. This opens up interesting possibilities of non-standard cosmology,
including an accelerating universe with $w<-1$.

A point is that the coefficient of the time kinetic term $\dot{\pi}^2$ 
in the low energy effective action for the scalar excitation $\pi$ is
positive and of order unity. This means that there is no ghost in the
ghost condensation. This also implies that in ghost condensation there
is no problem analogous to the strong coupling issues which the massive
gravity theory and the DGP brane model are facing with. On the other
hand, the coefficient of the space kinetic term $(\nabla\pi)^2$ in the
action becomes positive when $w<-1$. (A usual non-ghost scalar field,
with $w>-1$, has a positive coefficient for the time kinetic term and a 
negative coefficient for the space kinetic term.) This does not
necessarily introduce instabilities since there is also a
higher-derivative space kinetic term $(\nabla^2\pi)^2/M^2$ with a
negative coefficient. Actually, as far as the positive coefficient of
$(\nabla\pi)^2$ is small enough, the higher-derivative space kinetic
term pushes the would-be unstable modes outside the cosmological horizon
so that any instabilities do not show up. This situation is realized if 
the violation of the null energy condition is not too large since the
positive coefficient of $(\nabla\pi)^2$ is proportional to the amount of
violation of the null energy condition~\cite{Creminelli:2006xe}.

In the present paper we investigate the classical dynamics of cosmology
in ghost condensation with softly broken shift symmetry in more
detail. Throughout this paper we shall adopt a $4D$ covariant action
explained in Sec.~\ref{sec:4Daction}. In Sec.~\ref{sec:FRWbackground} we
show that it is always possible to find a form of the potential in the
Higgs sector Lagrangian which realizes an arbitrary FRW cosmological
history ($H(t)$, $\rho(t)$). A similar result is known to hold also for
a conventional scalar field with a potential, but in this case it is
impossible to violate the null energy condition without a ghost. This is
the origin of the folklore that the phantom ($w<-1$) cosmology requires
a ghost, which is correct for ordinary scalars. On the other hand, for
the ghost condensate, the null energy condition can be violated without
introducing ghosts or any other instabilities as far as the violation is
weak enough~\cite{Creminelli:2006xe}. Thus, the folklore is not correct
in ghost condensation. It is probably worth stressing here again that
the low energy effective field theory of ghost condensation is
completely determined by the symmetry breaking pattern and does not
include any ghosts.

After showing the reconstruction method in Sec.~\ref{sec:FRWbackground},
we investigate cosmological perturbation around the FRW background in
Sec.~\ref{sec:perturbation}. The resulting evolution equation is
summarized in subsection~\ref{subsec:results-perturbation} and can be
used to test the theory by observational data of e.g. cosmic microwave
background anisotropy, weak gravitational lensing and galaxy clustering
and so on.

\section{$4D$ covariant action}
\label{sec:4Daction}

Since the structure of the low energy effective field theory of ghost
condensation is determined by the symmetry breaking pattern, it is
not compulsory to consider a $4D$ covariant description for the ghost 
condensation as far as physics in the Higgs phase is
concerned. Nonetheless, it is instructive and sometimes convenient to
have such a description. In this paper we shall adopt the $4D$ covariant
description and extend the low energy field theory developed in
ref.~\cite{Arkani-Hamed:2003uy} to a general FRW background driven by not
only the ghost condensate itself but also other cosmological
fluids~\footnote{
See Sec.~3.3 of Ref.~\cite{Creminelli:2006xe} for the low energy
effective field theory in a FRW background driven solely by the ghost
condensate itself. In the present paper we include general gravitational
sources as well as the ghost condensate since the inclusion of those 
sources is essential for the test of the theory by e.g. cosmic microwave
background anisotropy, weak gravitational lensing, galaxy clustering and
so on.}.
In this language, we start with a $4D$ covariant action principle for a
scalar field, and the ghost condensation is realized as a background
with a non-vanishing derivative of the scalar field which does not
vanish even in Minkowski or de Sitter spacetime. Then the low energy
effective field theory is obtained by expanding the covariant action
around this background. Needless to say, the two approaches, one based
on the symmetry breaking pattern and the other based on the $4D$
covariant action, completely agree when applied to physics within the
regime of validity of the effective field theory.

The leading $4D$ covariant action for ghost condensation is given
by
%
\begin{equation}
 I = \int d^4x\sqrt{-g}
  \left[ \frac{M^4}{8}\Sigma^2 - M^4V(\phi)
			- \frac{\alpha_1}{2M^2}(\Box\phi)^2
			- \frac{\alpha_2}{2M^2}
			(\nabla^{\mu}\nabla^{\nu}\phi)
			(\nabla_{\mu}\nabla_{\nu}\phi)
	 \right],
  \label{eqn:4Daction}
\end{equation}
where 
%
\begin{equation}
 \Sigma \equiv -\frac{\partial^{\mu}\phi\partial_{\mu}\phi}{M^4} - 1,
\end{equation}
$M$ is the symmetry breaking scale giving the cutoff scale of the low
energy effective theory, and the sign convention for the metric is
($-+++$). Note that the potential $M^4V(\phi)$ is included to take into
account the soft breaking of the shift
symmetry~\cite{Creminelli:2006xe}: $V(\phi)$ is assumed to 
depend on $\phi$ very weakly. When $V(\phi)$ is a constant, the shift
symmetry is exact. The corresponding stress-energy tensor is easily
calculated as 
%
\begin{eqnarray}
 T^{(\phi)}_{\mu\nu} & = &
  \frac{1}{2}\Sigma\partial_{\mu}\phi\partial_{\nu}\phi 
  - \frac{\alpha_1+\alpha_2}{M^2}
  \left[\partial_{\mu}(\Box\phi)\partial_{\nu}\phi
   + \partial_{\mu}\phi\partial_{\nu}(\Box\phi)\right] \nonumber\\
 & &
  + \frac{\alpha_2}{M^2}
  \left\{
  \nabla^{\rho}
  \left[ (\nabla_{\mu}\nabla_{\nu}\phi)\partial_{\rho}\phi\right]
  - (R_{\mu}^{\rho}\partial_{\nu}\phi+R_{\nu}^{\rho}\partial_{\mu}\phi)
  \partial_{\rho}\phi \right\}  \nonumber\\
 & & + \left[ \frac{M^4}{8}\Sigma^2 - M^4V(\phi)
     +\frac{\alpha_1}{2M^2}(\Box\phi)^2
     +\frac{\alpha_1}{M^2}\partial^{\rho}(\Box\phi)\partial_{\rho}\phi
     \right.\nonumber\\
 & & \left.
     -\frac{\alpha_2}{2M^2}(\nabla^{\rho}\nabla^{\sigma}\phi)
     (\nabla_{\rho}\nabla_{\sigma}\phi)\right]g_{\mu\nu},
\end{eqnarray}
and the equation of motion for $\phi$ is 
%
\begin{equation}
 \frac{1}{2}\nabla^{\mu}(\Sigma\partial_{\mu}\phi) - M^4V'(\phi)
  - \frac{\alpha_1+\alpha_2}{M^2}\Box^2\phi 
  - \frac{\alpha_2}{M^2}
  \left(R^{\mu\nu}\nabla_{\mu}\nabla_{\nu}\phi
   +\frac{1}{2}\partial^{\mu}R\partial_{\mu}\phi \right)
  = 0. \label{eqn:EOM-phi}
\end{equation}
The Einstein equation is 
%
\begin{equation}
 M_{\rm Pl}^2G_{\mu}^{\nu} = T^{(\phi)\nu}_{\mu} + T_{\mu}^{\nu},
\end{equation}
where $T_{\mu}^{\nu}$ is the stress energy tensor of the other
gravitational sources satisfying the conservation equation 
$\nabla_{\nu}T_{\mu}^{\nu}=0$. As a consistency check, it is easy to
confirm that 
%
\begin{equation}
 \nabla_{\nu}T^{(\phi)\nu}_{\mu} = E^{(\phi)}\partial_{\mu}\phi, 
  \label{eqn:conservation-phi}
\end{equation}
where $E^{(\phi)}$ is the left hand side of (\ref{eqn:EOM-phi}). Thus,
the stress-energy tensor of $\phi$ satisfies the conservation equation 
$\nabla_{\nu}T^{(\phi)\nu}_{\mu}=0$, provided that the equation of
motion $E^{(\phi)}=0$ is satisfied. On the other hand, if the Einstein
equation is satisfied and if $\partial_{\mu}\phi$ is non-vanishing then
the equation of motion $E^{(\phi)}=0$ follows. Throughout this paper we
shall use these expressions for the stress-energy tensor and the
equation of motion.

If we set $\alpha_1=\alpha_2=0$ then the model is reduced to a kind of 
k-inflation~\cite{Armendariz-Picon:1999rj} or
k-essence~\cite{Chiba:1999ka,Armendariz-Picon:2000dh,Armendariz-Picon:2000ah}.
However, in this case there is no modification of gravity in Minkowski 
or de Sitter spacetime. Moreover, with $\alpha_1=\alpha_2=0$, the
attractor $\Sigma=0$ is unstable against inhomogeneous perturbations. In
the presence of the terms proportional to $\alpha_1$ and $\alpha_2$
($\alpha_1+\alpha_2>0$), the attractor is stable against small
perturbations and gravity is modified in the linearized level even in
Minkowski and de Sitter backgrounds. The relation between the
k-inflation and the ghost condensation is in some sense similar to that
between the usual potentially-dominated inflation and the Higgs
mechanism~\cite{Mukohyama:2005qd}.

\section{FRW background}
\label{sec:FRWbackground}

Ghost condensation provides the simplest Higgs phase of gravity in which
there is only one Nambu-Goldstone boson associated with spontaneous
Lorentz breaking. Note again that the dynamics of the Higgs phase of
gravity in ghost condensation has nothing to do with ghosts. Indeed, 
there is no ghost within the regime of validity of the effective field
theory. Moreover, the Higgs phase of gravity has universal low energy
description determined solely by the symmetry breaking pattern.

In this section we consider a flat FRW ansatz 
%
\begin{eqnarray}
 g_{\mu\nu}dx^{\mu}dx^{\nu} & = & -dt^2 + a(t)^2\delta_{ij}dx^idx^j,
  \nonumber\\
 \phi & = & \phi(t),
\end{eqnarray}
where $i=1,2,3$, and analyze the dynamics of the homogeneous, isotropic
universe. Linear perturbation around this background will be considered
in the next section. With this ansatz, the stress energy tensor for the
field $\phi$ is 
%
\begin{equation}
 T^{(\phi)\nu}_{\mu} = \left(\begin{array}{cccc} 
        -\rho_{\phi} & 0 & 0 & 0 \\
        0 & p_{\phi} & 0 & 0 \\
        0 & 0 & p_{\phi} & 0 \\
        0 & 0 & 0 & p_{\phi}
        \end{array}\right),
\end{equation}
where
%
\begin{eqnarray}
 \rho_{\phi} & = & \frac{1}{8}M^4(4+3\Sigma)\Sigma + M^4V \nonumber\\
& &  + \alpha M^2 
  \left[ \frac{3}{2}(1+\Sigma)(2\partial_tH-3H^2)
   + \frac{1}{2}\partial_t^2\Sigma
   - \frac{3}{8}\frac{(\partial_t\Sigma)^2}{1+\Sigma}
  \right] \nonumber\\
 & & +\beta M^2
  \left[ -3(1+\Sigma)\partial_tH+\frac{3}{2}H\partial_t\Sigma
  \right],\nonumber\\
 \rho_{\phi} + p_{\phi} & = & 
  \frac{M^4}{2}(1+\Sigma)\Sigma \nonumber\\
 & &  + \alpha M^2 
  \left[ 6(1+\Sigma)\partial_tH+(\partial_t^2\Sigma+3H\partial_t\Sigma)
   -\frac{1}{2}\frac{(\partial_t\Sigma)^2}{1+\Sigma}\right] \nonumber\\
 & &   - \beta M^2
  \left[ (1+\Sigma)(3H^2+5\partial_tH)
   + \frac{1}{2}(\partial_t^2\Sigma+H\partial_t\Sigma)
  \right],
  \label{eqn:rho-p-phi}
\end{eqnarray}
and
%
\begin{equation}
 \alpha \equiv \alpha_1+\alpha_2, \quad \beta\equiv\alpha_2.
\end{equation}
Here, we have expressed $\partial_t\phi$, $\partial_t^2\phi$ and
$\partial_t^3\phi$ in terms of $\Sigma=(\partial_t\phi)^2/M^4-1$ and its 
derivatives. The Einstein equation is
%
\begin{eqnarray}
 3M_{\rm Pl}^2H^2 & = & \rho_{\phi} + \rho, \nonumber\\
 2M_{\rm Pl}^2\partial_tH & = & -(\rho_{\phi}+p_{\phi}) - (\rho+p),
  \label{eqn:Einstei-eq-FRW}
\end{eqnarray}
where $H=\partial_ta/a$, and $\rho$ and $p$ are the energy density and
the pressure of the other gravitational sources. Following the usual
convention we call the first equation the Friedmann equation and the
second the dynamical equation. Note that, because of the identity 
(\ref{eqn:conservation-phi}) and the Bianchi identity, the equation of 
motion for $\phi$ automatically follows from (\ref{eqn:Einstei-eq-FRW}),
provided that the conservation equation for $\rho$ and $p$ holds: 
%
\begin{equation}
 \partial_t\rho + 3H(\rho+p) = 0.
\end{equation}

\subsection{Low energy expansion}
\label{subsec:lowE-expansion-background}

At this point, one might think that the expressions in
(\ref{eqn:rho-p-phi}) are too complicated to extract the physical
picture of cosmology in the Higgs phase of gravity. Even if one could
somehow manage to do so with brute force for the FRW background, it
might be too optimistic to expect that the same approach works for 
linear perturbation around the FRW background. So let us go back and
reconsider the meaning of the covariant action (\ref{eqn:4Daction}). 
This is not a full action including a UV completion but just a leading
action suitable to describe physics sufficiently below the cutoff scale
$M$ around backgrounds with $\Sigma\simeq 0$. If the Hubble expansion
rate and/or the stress-energy tensor of the field $\phi$ become close to
or above unity in the unit of $M$ then the low energy description is
invalidated and we need a UV 
completion. Needless to say, the same criterion applies to the other
approach based on the symmetry breaking pattern, and the two
approaches agree in the regime of validity of the low energy effective
theory. Therefore, all we can and should trust is what is obtained below
the cutoff scale. In other words, we assume the existence of a good UV
completion but never use its properties. For this reason, we can and
should ignore terms irrelevant at low energies compared with the cutoff
scale $M$.

To be systematic, we adopt a low energy expansion by introducing small
dimensionless parameters $\epsilon_i$ ($i=0,1,2,3$) as
%
\begin{equation}
 \frac{M^2}{M_{\rm Pl}^2} = \epsilon_0, \quad
 \frac{H}{M} = \epsilon_1, \quad
 \frac{\rho}{M_{\rm Pl}^2M^2} = \epsilon_2^2, \quad
 \frac{p}{M_{\rm Pl}^2M^2} = O(\epsilon_2^2),
\end{equation}
and 
%
\begin{equation}
 {\rm Max}(|\Sigma|,|V|)=\epsilon_3, \quad
  \left( \Sigma = O(\epsilon_3), \quad 
   V = O(\epsilon_3) \right).
\end{equation}
Since the time scale for the change of $H$, $\Sigma$, $V$, $\rho$ and
$p$ is expected to be the cosmological time scale $1/H$, it also follows
that
%
\begin{equation}
 \frac{\partial_t^nH}{M^{n+1}} = O(\epsilon_1^{n+1}), \quad
 \frac{\partial_t^n\Sigma}{M^n} = O(\epsilon_3\epsilon_1^n), \quad
 \frac{\partial_t^n V}{M^n} = O(\epsilon_3\epsilon_1^n), 
\end{equation}
and that
%
\begin{equation}
 \frac{\partial_t^n\rho}{M_{\rm Pl}^2M^{n+2}}
  = O(\epsilon_2^2\epsilon_1^n), \quad
 \frac{\partial_t^n p}{M_{\rm Pl}^2M^{n+2}}
 = O(\epsilon_2^2\epsilon_1^n),
 \label{eqn:lowE-drho-dp-epsilon2}
\end{equation}
for $n=1,2,\cdots$. The latter condition
(\ref{eqn:lowE-drho-dp-epsilon2}) is not necessary for the present
purpose but will be used when we derive evolution equations for linear
perturbation around the FRW background in the next section. Unless
fine-tuned, consistency of this assignment with the Einstein equation 
(\ref{eqn:Einstei-eq-FRW}) requires that 
%
\begin{equation}
 \epsilon_1^2 \simeq 
  {\rm Max} (\epsilon_0\epsilon_3,\epsilon_2^2).
  \label{eqn:consistency-epsilons}
\end{equation}

It is in principle possible but not practical to perform the low energy
expansion with respect to all $\epsilon$'s, considering each
$\epsilon_i$ as independent small parameters. Of course, if one performs 
the low energy expansion up to sufficiently high order with respect to
all $\epsilon$'s then one can always reach the point where all relevant
terms (and probably many other irrelevant terms) are included. However,
this is not the most economical way to obtain results relevant to
particlar situations of physical interest. It is more economical and
convenient to suppose some rough relations among these small parameters
to reduce the number of independent small parameters. This also makes it
easier to extract physical picture out of complicated equations. Note
that those relations among $\epsilon_i$ must reflect the situations of
physical  interest.

For example, in ghost inflation~\cite{Arkani-Hamed:2003uz} we set
$\Sigma=V=0$ and include another field, say $\psi$, to end the inflation
and to reheat the universe a la hybrid inflation. In this case
$\epsilon_3$ vanishes. Moreover, since the stress energy tensor is
dominated by the other field $\psi$, the consistency relation
(\ref{eqn:consistency-epsilons}) implies that 
$\epsilon_1\simeq\epsilon_2$. In this way, in
Ref.~\cite{Arkani-Hamed:2003uz} we considered $\epsilon_0$ and 
$\epsilon_1$ ($\simeq\epsilon_2$) as two independent small parameters.

On the other hand, in the present paper we would like to consider
late time cosmology in which both ($\rho_{\phi}$, $p_{\phi}$) and 
($\rho$, $p$) may contribute to the background as gravitational
sources. For this reason and because of the consistency relation
(\ref{eqn:consistency-epsilons}), we suppose that there is a small
number $\epsilon$ such that 
%
\begin{equation}
 \epsilon_1^2 = O(\epsilon^2), \quad 
  \epsilon_0\epsilon_3 = O(\epsilon^2), \quad
  \epsilon_2^2 = O(\epsilon^2), 
  \quad
  \left( \mbox{e.g. }
   \epsilon \equiv {\rm Max}(\sqrt{\epsilon_0\epsilon_3},\epsilon_2)
  \right).
  \label{eqn:epsilon0-3}
\end{equation}
We expect (and will actually show) that in the dispersion relation for
the excitation of ghost condensation, $M^2/M_{\rm Pl}^2$ ($=\epsilon_0$) 
and $\Sigma$ ($=\epsilon_3$) additively contribute to the coefficient of
the momentum squared. (See the expression for $C_0$ in (\ref{eqn:C-def})
below.) We would like to consider situations in which they can
independently become relevant since they may become different sources of
possible instabilities and interesting physical
effects~\cite{Creminelli:2006xe}. Thus, we refine the second relation in
(\ref{eqn:epsilon0-3}) to 
%
\begin{equation}
 \epsilon_0 = O(\epsilon), \quad \epsilon_3=O(\epsilon),\quad
  \left( \mbox{e.g. }
   \epsilon \equiv {\rm Max}
   (\epsilon_0,\epsilon_2,\epsilon_3)
       \right).
\end{equation}
In summary we can introduce just one small parameter $\epsilon$ and
suppose that 
%
\begin{equation}
 \frac{M^2}{M_{\rm Pl}^2} = O(\epsilon), \quad
 \frac{\partial_t^nH}{M^{n+1}} = O(\epsilon^{n+1}), \quad
 \frac{\partial_t^n\Sigma}{M^n} = O(\epsilon^{n+1}), \quad
 \frac{\partial_t^n V}{M^n} = O(\epsilon^{n+1}), 
 \label{eqn:lowE-background}
\end{equation}
and
%
\begin{equation}
 \frac{\partial_t^n\rho}{M^{n+4}} = O(\epsilon^{n+1}), \quad
 \frac{\partial_t^n p}{M^{n+4}} = O(\epsilon^{n+1}),
 \label{eqn:lowE-drho-dp}
\end{equation}
where $n=0,1,2,\cdots$. Again, the latter condition
(\ref{eqn:lowE-drho-dp}) is not necessary for the present purpose but
will be used when we derive evolution equations for linear perturbation
around the FRW background in the next section.

The low energy expansion not only enables us to compute various
quantities in a systematic way but also helps us avoid picking up 
spurious modes associated with the higher derivative terms. In the $4D$
covariant action (\ref{eqn:4Daction}) the terms proportional to
$\alpha_1$ and $\alpha_2$ include the square of the second time
derivative of the field $\phi$. Therefore, if we take its face value
then the equation of motion for $\phi$ includes up to the fourth order
time derivatives and there is in principle freedom to specify $\phi$,
$\partial_t\phi$, $\partial_t^2\phi$ and $\partial_t^3\phi$ as an
initial condition. However, as the scaling analysis in
ref.~\cite{Arkani-Hamed:2003uy} shows, the time derivatives higher than
the second order in the equation of motion are irrelevant at energies
sufficiently below the cutoff $M$, at least in Minkowski background. In
other words, extra modes associated with those higher time derivatives
have frequencies above the cutoff scale $M$ and is outside the regime of 
validity of the effective field theory. For this reason, those extra
modes are spurious and must be dropped out from physical spectrum of the
low energy theory. In the expanding background it is not a priori
completely clear how to drop the spurious modes while maintaining all
physical modes. By adopting the low energy expansion this can be done in
a systematic way.

\subsection{Reconstructing the potential from H(t)}
\label{subsec:reconstruct-V}

As shown in Ref.~\cite{Creminelli:2006xe}, many non-standard cosmology, 
including the phantom ($w<-1$) cosmology, can be realized in the
framework of ghost condensation without introducing ghosts or any other
instabilities. The purpose of this subsection is to show that, given an
arbitrary history of the Hubble expansion rate $H(t)$ and gravitational
sources $\rho(t)$ and $p(t)$, it is indeed possible to find a form of
the potential $V(\phi)$ which realizes $H(t)$ as a solution to the
Einstein equation (\ref{eqn:Einstei-eq-FRW}), provided that the
conservation equation $\partial_t\rho+3H(\rho+p)=0$ is satisfied. As is
well known, a similar result holds also for a conventional scalar field
with a potential: one can almost always find a form of the potential for
a given history of the Hubble expansion rate. However, for a
conventional scalar field, it is impossible to violate the null energy
condition without a ghost. This is the origin of the folklore that the
phantom ($w<-1$) cosmology requires a ghost, which is correct for
ordinary scalars. On the other hand, for the ghost condensate, the null
energy condition can be violated without introducing ghosts or any other
instabilities as far as the violation is weak enough. Thus, the folklore
is not correct in ghost condensation. It is probably worth stressing
here again that the low energy effective field theory of ghost
condensation is completely determined by the symmetry breaking pattern
and does not include any ghosts.

For the reconstruction of the potential $V(\phi)$ from the expansion
history $H(t)$ and the energy density $\rho(t)$ and pressure $p(t)$ of
known gravitational sources, we take advantage of the low energy
expansion introduced in the previous subsection. We expand $V$ and
$\Sigma$ with respect to $\epsilon$: 
%
\begin{eqnarray}
 V & = & V_0 + V_1 + \cdots, \quad 
  V_n = O(\epsilon^{n+1}), \nonumber\\
 \Sigma & = & \Sigma_0 + \Sigma_1 + \cdots, \quad
  \Sigma_n = O(\epsilon^{n+1}). 
\end{eqnarray}
Note that $\phi$ is expressed in terms of $\Sigma_n$ as
%
\begin{equation}
 \phi = M^2 \int \sqrt{1+\Sigma}dt = 
  M^2
  \int\left[1+\frac{1}{2}\Sigma_0
       +O(\epsilon^2)\right]dt.
  \label{eqn:phi-Sigma}
\end{equation}

The Friedmann and dynamical equations (\ref{eqn:Einstei-eq-FRW}) in the
lowest order in $\epsilon$ are easily solved with respect to $\Sigma_0$
and $V_0$ as 
%
\begin{eqnarray}
 \Sigma_0 & = & -\frac{2}{M^4}
  \left[2M_{\rm Pl}^2\partial_tH+(\rho+p)\right],
  \nonumber\\
 V_0 & = & \frac{1}{M^4}
  \left[M_{\rm Pl}^2(2\partial_tH+3H^2)+p\right],
  \label{eqn:sol-Sigma0-V0}
\end{eqnarray}
The relation (\ref{eqn:phi-Sigma}) in the lowest order in $\epsilon$ is 
%
\begin{equation}
 \phi = M^2 (t-t_0) + M\cdot O(\epsilon), 
\end{equation}
where $t_0$ is a constant. Thus, for a given history ($H(t)$, $\rho(t)$,
$p(t)$) satisfying $\partial_t\rho+3H(\rho+p)=0$, the potential $V$ is
expressed in terms of $\phi$ as  
%
\begin{equation}
 V(\phi) = 
  \frac{1}{M^4}\left[M_{\rm Pl}^2(2\partial_tH+3H^2)+p\right]_{t=t_0+\phi/M^2}
  +O(\epsilon^2).
  \label{eqn:V-phi-leading}
\end{equation}

It is also easy to obtain the correction to the lowest order potential. 
In the next-to-the-leading order the Friedmann and dynamical equations 
(\ref{eqn:Einstei-eq-FRW}) are solved with respect to $\Sigma_1$ and
$V_1$ as 
%
\begin{eqnarray}
 \Sigma_1 & = & -\Sigma_0^2 - \frac{12\alpha}{M^2}\partial_tH
  +\frac{2\beta}{M^2}(5\partial_tH+3H^2), 
  \nonumber\\
 V_1 & = & \frac{1}{8}\Sigma_0^2
  +\frac{3\alpha-2\beta}{2M^2}(2\partial_tH+3H^2). 
  \label{eqn:sol-Sigma1-V1}
\end{eqnarray}
The relation (\ref{eqn:phi-Sigma}) in this order is
%
\begin{equation}
 \phi = M^2 (t-t_0) 
  + \frac{M^2}{2}\int_{t_0}^t \Sigma_0(t')dt'
       + M\cdot O(\epsilon^2), 
\end{equation}
where $t_0$ is again a constant. Thus, for a given history ($H(t)$,
$\rho(t)$, $p(t)$) satisfying $\partial_t\rho+3H(\rho+p)=0$, the
potential $V$ is expressed in terms of $\phi$ as  
%
\begin{equation}
 V(\phi) =
  \left[
   V_0 + \Delta t\cdot\partial_t V_0 + V_1\right]_{t=t_0+\phi/M^2}
  + O(\epsilon^3),
\end{equation}
where 
%
\begin{equation}
 \Delta t =  -\frac{1}{2}\int_{t_0}^t\Sigma_0(t')dt'
\end{equation}
and $\Sigma_0$, $V_0$ and $V_1$ are given in (\ref{eqn:sol-Sigma0-V0})
and (\ref{eqn:sol-Sigma1-V1}).

\section{Cosmological perturbation}
\label{sec:perturbation}

In this section we derive evolution equations for cosmological
perturbation around the FRW background.

We consider a scalar-type cosmological perturbation in the longitudinal
gauge,
%
\begin{equation}
 ds^2 = -(1+2\Psi Y)dt^2
  + a^2(1+2\Phi Y)d{\bf x}^2,
\end{equation}
and a general stress-energy tensor of the form 
%
\begin{eqnarray}
 T^t_t & = & - \rho
  - \left[\rho\Delta-3H(\rho+p)Va/\sqrt{{\bf k}^2}\right]Y, \nonumber\\
 T^t_i & = & (\rho+p)Va Y_i, \quad
 T^i_t = -\frac{(\rho+p)V Y^i}{a}, \nonumber\\
 T^i_j & = & p
  + \left\{ p\Gamma+ c_s^2[\rho\Delta
     -3H(\rho+p)Va/\sqrt{{\bf k}^2}]\right\}Y\delta^i_j
  + p\Pi Y^i_j,
\end{eqnarray}
where $Y$'s are harmonics on the $3$-space defined as
%
\begin{eqnarray}
 Y & = & e^{i{\bf k}\cdot{\bf x}}, \nonumber\\
 Y_i & = & -\frac{1}{\sqrt{{\bf k}^2}}\partial_i Y, \quad
 Y^i = \delta^{ij}Y_j, \nonumber\\
 Y_{ij} & = & \frac{1}{{\bf k}^2}\partial_i\partial_jY
  +\frac{1}{3}Y\delta_{ij}, \quad
 Y^i_j = \delta^{ik}Y_{kj}.
\end{eqnarray}
Here, $\rho$ and $p$ are the unperturbed energy density and pressure,
and ($\Delta$, $V$, $\Gamma$, $\Pi$) represent the gauge-invariant
perturbation of the stress-energy tensor. Physical meaning of each
component is as follows~\cite{Kodama:1985bj}: $\Delta$ is the density 
contrast in the slicing that the fluid velocity is orthogonal to
constant time hypersurfaces, $V$ is the fluid velocity relative to the
observers normal to constant time hypersurfaces, $\Gamma$ is the entropy
perturbation and $\Pi$ is anisotropic stress. They satisfy the perturbed
conservation equation  
%
\begin{eqnarray}
 \partial_t(\rho\Delta)+3H\rho\Delta
  + \left(\frac{{\bf k}^2}{a^2}-3\partial_t H\right)
  \frac{a}{\sqrt{{\bf k}^2}}(\rho+p)V + 2Hp\Pi & & \nonumber\\
  + 3(\rho+p)(\partial_t\Phi-H\Psi) & = & 0, \nonumber\\
 \partial_t\left[(\rho+p)V\right]+(4+3c_s^2)H(\rho+p)V
  -\frac{\sqrt{{\bf k}^2}}{a}
  \left[ p\Gamma + c_s^2\rho\Delta+(\rho+p)\Psi-\frac{2}{3}p\Pi
  \right] & = & 0.\nonumber\\
\end{eqnarray}
When $\rho$ and $\rho+p$ are non-vanishing, these equations are
rewritten as
%
\begin{eqnarray}
 \partial_t\Delta - 3wH\Delta + (1+w)
  \left[\left(\frac{{\bf k}^2}{a^2}-3\partial_tH\right)
   \frac{a}{\sqrt{{\bf k}^2}}V+3(\partial_t\Phi-H\Psi)\right]
  +2wH\Pi & = & 0,\nonumber\\
 \partial_t V + HV - \frac{\sqrt{{\bf k}^2}}{a}
  \left[\frac{c_s^2}{1+w}\Delta+\frac{w}{1+w}
   \left(\Gamma-\frac{2}{3}\Pi\right)+\Psi\right] & = & 0, 
  \label{eqn:conservation-perturbation}
  \nonumber\\
\end{eqnarray}
where $w=p/\rho$ and $c_s^2=\partial_t p/\partial_t\rho$.

As for the metric perturbation, for convenience, we decompose
($\Phi$,$\Psi$) into the standard, general relativity (GR) part
($\Phi_{GR}$,$\Psi_{GR}$) and the modification part
($\Phi_{mod}$,$\Psi_{mod}$) as follows. 
%
\begin{eqnarray}
 \Phi & = & \Phi_{GR} + \Phi_{mod}, \nonumber\\
 \Psi & = & \Psi_{GR} + \Psi_{mod},
  \label{eqn:GR-plus-mod}
\end{eqnarray}
where the GR part is given by 
%
\begin{eqnarray}
 \frac{{\bf k}^2}{a^2}\Phi_{GR} = 
  \frac{\rho\Delta}{2M_{\rm Pl}^2}, \nonumber\\
 \frac{{\bf k}^2}{a^2}(\Psi_{GR}+\Phi_{GR}) = 
  -\frac{p\Pi}{M_{\rm Pl}^2}. 
  \label{eqn:GRpart-def}
\end{eqnarray}

In the following we shall derive evolution equation of the modification
part of the metric perturbation ($\Phi_{mod}$, $\Psi_{mod}$). For
readers who are interested in application of the formalism,
subsection~\ref{subsec:results-perturbation} summarizes resulting 
evolution equations of the cosmological perturbation in the Higgs phase
of gravity. In subsection~\ref{subsec:lowE-expansion-perturbation} we
introduce a systematic low energy expansion applicable to linear
perturbation by extending the low energy expansion for the background
given in subsection~\ref{subsec:lowE-expansion-background}. 
Subsection~\ref{subsec:master-eq} includes the hardest part of
calculations in this paper: we derive a single master equation governing 
the modification of gravity without ignoring higher time
derivatives. The master equation is a fourth-order ordinary differential 
equation for each comoving momentum. However, in 
subsection~\ref{subsec:reduction} we see that the fourth and third order 
derivative terms are irrelevant and reduce the master equation to a set 
of two first-order ordinary differential equations for each comoving 
momentum. At the same time, we remove an apparent singularity in the
master equation. In subsection~\ref{subsec:results-perturbation} we
summarize the results of this section in a way which is directly
applicable to actual problems. In Appendix~\ref{app:example}, as a 
simple application of the formula summarized in
subsection~\ref{subsec:results-perturbation}, we consider Minkowski and
de Sitter backgrounds and derive the results in
ref.~\cite{Arkani-Hamed:2003uy} for modification of gravity in those
backgrounds. Readers who are interested in application of the formalism
may go directly to subsections~\ref{subsec:results-perturbation}.

\subsection{Low energy expansion}
\label{subsec:lowE-expansion-perturbation}

We now extend the low energy expansion developed in
subsection~\ref{subsec:lowE-expansion-background} to linear perturbation
around the FRW background. The assignment for the FRW background is
summarized in (\ref{eqn:lowE-background}) and (\ref{eqn:lowE-drho-dp}).

For perturbation, besides the smallness of the perturbation itself
controlling the validity of the linear approximation, there is one
additional small parameter $\epsilon_4$ given by  
%
\begin{equation}
 \frac{1}{M}\frac{\sqrt{{\bf k}^2}}{a} = \epsilon_4. 
\end{equation}
The low energy effective theory is valid only if $\epsilon_4$ is small
enough. Supposing that $\Delta$, $V$, $\Gamma$ and $\Pi$ may become
comparable, the perturbed conservation equation
(\ref{eqn:conservation-perturbation}) implies that
%
\begin{equation}
 \frac{\partial_t}{M}(\Delta,V,\Gamma,\Pi)
  \sim (\Delta,V,\Gamma,\Pi) \cdot
  {\rm Max}(\epsilon_1,\epsilon_4),
\end{equation}
unless the modification part ($\Phi_{mod}$, $\Psi_{mod}$) dominates over
the GR part ($\Phi_{GR}$, $\Psi_{GR}$). As shown already in
ref.~\cite{Arkani-Hamed:2003uy}, the dispersion relation for excitation
of ghost condensation in Minkowski background is 
$\omega^2\simeq 
\alpha {\bf k}^4/M^2 -\alpha{\bf k}^2M^2/2M_{\rm Pl}^2$. 
The first term in the right hand side is $\sim \epsilon_4^4M^2$ and the
second term is $\sim \epsilon\epsilon_4^2M^2$. 
In this paper we would like to generalize this dispersion relation in
Minkowski spacetime to an evolution equation in the FRW background,
taking both terms into account. For this purpose we suppose that these
two terms may become comparable. In other words, we suppose that there
is a small parameter $\tilde{\epsilon}$ such that
%
\begin{equation}
 \epsilon = O(\tilde{\epsilon}^2), \quad
  \epsilon_4 = O(\tilde{\epsilon}), \quad
  \left(\mbox{e.g. }
   \tilde{\epsilon} \equiv {\rm Max}
   (\sqrt{|\epsilon|},\epsilon_4)
  \right).
\end{equation}

In summary it is supposed that there is a small parameter
$\tilde{\epsilon}$ such that
%
\begin{eqnarray}
 \frac{M^2}{M_{\rm Pl}^2} & = & O(\tilde{\epsilon}^2), \nonumber\\
 \frac{\partial_t^nH}{M^{n+1}} & = & O(\tilde{\epsilon}^{2(n+1)}), 
  \quad
 \frac{\partial_t^n\Sigma}{M^n} = O(\tilde{\epsilon}^{2(n+1)}), \quad 
 \frac{\partial_t^n V}{M^n} = O(\tilde{\epsilon}^{2(n+1)}), \nonumber\\
 \frac{\partial_t^n\rho}{M^{n+4}} & = & O(\tilde{\epsilon}^{2(n+1)}), 
  \quad
 \frac{\partial_t^n p}{M^{n+4}} = O(\tilde{\epsilon}^{2(n+1)}),
\end{eqnarray}
and
%
\begin{eqnarray}
 \frac{1}{M}\frac{\sqrt{{\bf k}^2}}{a} & = & O(\tilde{\epsilon}),
  \nonumber\\
 \left(\frac{\partial_t}{M}\right)^n(\Delta,V,\Gamma,\Pi)
  & \sim & O(\tilde{\epsilon}^n) \cdot (\Delta,V,\Gamma,\Pi),
\end{eqnarray}
where $n=0,1,2,\cdots$.

\subsection{Master equation}
\label{subsec:master-eq}

In the previous subsection we developed a low energy expansion for the
FRW background and linear perturbation around it. Armed with this, we
can roughly estimate how small or large a term in equations should be, as
far as ($\Phi_{GR}$, $\Psi_{GR}$), ($\Delta$, $V$, $\Gamma$, $\Pi$) and
their time derivatives are concerned. On the other hand, we do not know
a priori how small or large the time derivatives of the modification 
part ($\Phi_{mod}$, $\Psi_{mod}$) are. This can be seen only after we
derive evolution equation of these quantities. In this section we derive
a single fourth order master equation for the modification part, using
the low energy expansion but not assuming the smallness of the time
derivatives of ($\Phi_{mod}$, $\Psi_{mod}$).

Let us introduce two variables $\tilde{\Phi}_{1,2}$ by 
%
\begin{eqnarray}
 \Phi_{mod} & = & \tilde{\Phi}_1 + c\tilde{\Phi}_2, \nonumber\\
 \Psi_{mod} & = & - \tilde{\Phi}_1 - (1+c)\tilde{\Phi}_2,
  \label{eqn:def-Phi1-Phi2}
\end{eqnarray}
where $c$ is a constant. We shall see below that the source terms for
equations governing $\Phi_{1,2}$ scale like $\propto M^2/M_{\rm Pl}^2$
in the $M/M_{\rm Pl}\to 0$ limit. Thus the standard GR result is
recovered in the $M/M_{\rm Pl}\to 0$ limit. The variable
$\tilde{\Phi}_2$ ($\propto\Phi_{mod}+\Psi_{mod}$) was defined so that it
vanishes when $\beta=0$. The definition of the variable $\tilde{\Phi}_1$
is not yet fixed because of the unfixed constant $c$. The value of the
constant $c$ will later be determined so that the master equation for 
$\tilde{\Phi}_1$ including matter source terms is simplified.

As for the field $\phi$ responsible for ghost condensation, we expand it
up to the linear order as 
%
\begin{equation}
 \phi = M^2 \left[\int \sqrt{1+\Sigma}dt + \pi Y\right].
\end{equation}
We now have three unknown variables $\tilde{\Phi}_1$, $\tilde{\Phi}_2$
and $\pi$ other than matter variables ($\Delta$, $V$, $\Gamma$,
$\Pi$). Our task now is to obtain the evolution equation for those three
variables sourced by these matter variables.

In cosmology with an ordinary scalar field the standard strategy to
analyze linear perturbation in the longitudinal gauge is as follows: (i)
to eliminate one of the two metric variables ($\Phi$, $\Psi$), say
$\Psi$, by using the traceless part of ($ij$)-components of the
linearized Einstein equation; (ii) to eliminate the linear perturbation
of the scalar field by using the ($0i$)-components of the linearized
Einstein equation; and (iii) to obtain a single master equation for the
remaining metric variable, say $\Phi$, from the ($00$)-component of the
linearized Einstein equation. (The linearized equation of motion of the
scalar field and the trace part of ($ij$)-components of the linearized
Einstein equation are automatically satisfied because of the Bianchi
identity.) The steps (i) and (ii) involve simple algebraic equations
for variables being eliminated. As a result the master equation is a
second-order ordinary differential equation for each comoving
momentum. This is consistent with the fact that, in the absence of other
gravitational sources, the scalar sector includes only one propagating
degree of freedom, i.e. excitation of the scalar field. If there are
other gravitational sources then the master equation includes source
terms due to those gravitational sources.

We can use the same strategy to analyze the linear perturbation with
ghost condensation although the result will be drastically
different. There are again three main steps. (To be precise, 
for the reason explained below, steps (ii) and (iii) are not separated
but actually mixed in the present case.) (i) First we eliminate
$\tilde{\Phi}_2$ by using the traceless part of ($ij$)-components of the
linearized Einstein equation. (ii) Next we eliminate $\pi$ by using the
($0i$)-components of the linearized Einstein equation. (iii) Finally we 
obtain a single master equation for the remaining metric variable
$\tilde{\Phi}_1$ from the ($00$)-component of the linearized Einstein
equation. (The equation of motion of $\pi$ and the trace part of
($ij$)-components of the linearized Einstein equation are automatically 
satisfied because of the Bianchi identity.) The step (i) involves just
an algebraic equation for $\tilde{\Phi}_2$ as in the standard
case. However, the step (ii) involves a second-order differential
equation for $\pi$ for each comoving momentum, contrary to the standard
case. For this reason the steps (ii) and (iii) are not separated but
actually mixed. After eliminating $\pi$ and its derivatives, the
resulting master equation is a fourth-order ordinary differential
equation for $\tilde{\Phi}_1$ for each comoving momentum. This is indeed
consistent with the fact that the action for $\phi$ includes the square
of the second derivative of $\phi$ and that the equation of motion for
$\pi$ involves up to forth-order derivatives at least formally. In the
master equation the matter variables ($\Delta$, $V$, $\Gamma$, $\Pi$)
appear as source terms.

We adopt the low energy expansion introduced in the previous subsection
to follow each step (i)-(iii). This in particular means that terms with
sufficiently higher time derivatives acted on the background or matter
variables can be ignored since they are considered as higher order in
the expansion with respect to $\tilde{\epsilon}$. On the other hand, the
time derivative acted on $\tilde{\Phi}_1$, $\tilde{\Phi}_2$ and $\pi$ is
not supposed to raise the order of the low energy expansion. This is
because we do not a priori know the time scale of the dynamics of these
three variables in the FRW background until the evolution equation of a
single master variable is obtained.

Following the steps (i)-(iii), we obtain the fourth-oder master equation
for $\tilde{\Phi}_1$: 
%
\begin{equation}
 C_4\partial_t^4\tilde{\Phi}_1 + C_3\partial_t^3\tilde{\Phi}_1 + \partial_t^2\tilde{\Phi}_1 +
  C_1\partial_t\tilde{\Phi}_1 + C_0\tilde{\Phi}_1 = \tilde{S}_1,
\end{equation}
where the coefficient of the second-order term is normalized to unity,
and other coefficients and the source term are given by 
%
\begin{eqnarray}
 C_4 & = & \frac{\alpha}{M^2} + \frac{O(\tilde{\epsilon}^2)}{M^2},
  \nonumber\\
 C_3 & = & \frac{\alpha}{M^2}
  \left[ 6H-\frac{M^2(\partial_t\Sigma+2H\Sigma)}
   {M^2\Sigma+2\alpha{\bf k}^2/a^2}\right]
  + \frac{O(\tilde{\epsilon}^4)}{M},\nonumber\\
 C_1 & = & 3H - 
  \frac{M^2(\partial_t\Sigma+2H\Sigma)}{M^2\Sigma+2\alpha{\bf k}^2/a^2} +
  M\cdot O(\tilde{\epsilon}^4),\nonumber\\
 C_0 & = & \frac{\alpha}{M^2}\frac{{\bf k}^4}{a^4} +
  \frac{1}{2}\left(\Sigma-\alpha\frac{M^2}{M_{\rm Pl}^2}
  \right)\frac{{\bf k}^2}{a^2} + 2H^2+\partial_t H \nonumber\\
 & & -\frac{\Sigma M^4}{4M_{\rm Pl}^2}
  -\frac{M^2H(\partial_t\Sigma+2H\Sigma)}
  {M^2\Sigma+2\alpha{\bf k}^2/a^2}
  + M^2\cdot O(\tilde{\epsilon}^6),
  \label{eqn:C-def}
\end{eqnarray}
and
%
\begin{eqnarray}
 M_{\rm Pl}^2\frac{{\bf k}^2}{a^2}\tilde{S}_1 & = & \frac{M^2}{M_{\rm Pl}^2}
  \left\{\frac{1}{8}
   \left( M^2\Sigma+2\alpha{\bf k}^2/a^2\right)
   \left[(1+3c_s^2)\rho\Delta+3p\Gamma\right] 
   \right.\nonumber\\
 & & \left.
   + (c-1)\beta
   \left[
    \frac{1}{2}\frac{{\bf k}^2}{a^2}(c_s^2\rho\Delta+p\Gamma)
    -\frac{1}{3}\frac{{\bf k}^2}{a^2}p\Pi
   \right]
   + c\beta \partial_t^2(p\Pi) \right\}
 \nonumber\\
 & & + M^6\cdot O(\tilde{\epsilon}^5).
\end{eqnarray}
On the other hand, $\tilde{\Phi}_2$ is expressed in terms of $\tilde{\Phi}_1$ and its
derivatives as
%
\begin{equation}
 \tilde{\Phi}_2 = \beta\left(D_3\partial_t^3\tilde{\Phi}_1 
  + D_2\partial_t^2\tilde{\Phi}_1 + D_1\partial_t\tilde{\Phi}_1 +
  D_0\tilde{\Phi}_1 + S_2\right),
\end{equation}
where
%
\begin{eqnarray}
 D_3 & = &
  \frac{4\alpha H}
  {M^2\left[M^2\Sigma+2\alpha{\bf k}^2/a^2\right]}
  + \frac{O(\tilde{\epsilon}^2)}{M^3},\nonumber\\
 D_2 & = & -\frac{2\alpha}{M^2[\alpha-(2c+1)\beta]}
  + \frac{O(\tilde{\epsilon}^2)}{M^2},\nonumber\\
 D_1 & = &  
  \frac{4H}{M^2\Sigma+2\alpha{\bf k}^2/a^2}
  + \frac{O(\tilde{\epsilon}^2)}{M},  \nonumber\\
 D_0 & = & 
  -\frac{2}{M^2}\frac{{\bf k}^2}{a^2} + \frac{M^2}{M_{\rm Pl}^2} 
  +\frac{4H^2}{M^2\Sigma+2\alpha{\bf k}^2/a^2}
  + O(\tilde{\epsilon}^4), \nonumber\\
 M_{\rm Pl}^2\frac{{\bf k}^2}{a^2}S_2 & = & 
   \frac{M^2}{2M_{\rm Pl}^2}(\rho\Delta+2p\Pi)
   + M^4\cdot O(\tilde{\epsilon}^3).
   \label{eqn:def-D-S2}
\end{eqnarray}
To obtain these expressions, we kept terms up to $O(\tilde{\epsilon}^4)$
in intermediate steps.

Having obtained the master equation, it is clear that 
%
\begin{equation}
 c = 1
\end{equation}
gives the best choice for the definition of $\tilde{\Phi}_1$. This
indeed simplifies the source term $\tilde{S}_1$ of the master
equation. Moreover, the term $(M^2/M_{\rm Pl}^2)\beta\partial_t^2(p\Pi)$
in $\tilde{S}_1$ can be absorbed by redefinition of the master variable:
it does not appear if we use $\Phi_1$ and $\Phi_2$ defined by 
%
\begin{eqnarray}
 \Phi_1 & \equiv & \tilde{\Phi}_1 
  + \beta \frac{a^2}{{\bf k}^2}\frac{p\Pi}{M_{\rm Pl}^2}, \nonumber\\
 \Phi_2 & \equiv & \tilde{\Phi}_2,
\end{eqnarray}
instead of $\tilde{\Phi}_1$ and $\tilde{\Phi}_2$. The master equation is
now written as
%
\begin{equation}
 C_4\partial_t^4\Phi_1 + C_3\partial_t^3\Phi_1 +
  \partial_t^2\Phi_1 + C_1\partial_t\Phi_1 + C_0\Phi_1 = S_1,
  \label{eqn:master-eq-4th}
\end{equation}
where $C_{4,3,1,0}$ are given by (\ref{eqn:C-def}) and the new source
term is 
%
\begin{equation}
 M_{\rm Pl}^2\frac{{\bf k}^2}{a^2}S_1 = 
  \frac{M^2}{8M_{\rm Pl}^2}
  \left( M^2\Sigma+2\alpha{\bf k}^2/a^2\right)
  \left[(1+3c_s^2)\rho\Delta+3p\Gamma\right] 
  + M^6\cdot O(\tilde{\epsilon}^5).
  \label{eqn:S1-def}
\end{equation}
The variable $\Phi_2$ is expressed in terms of $\Phi_1$ and matter
variables as
%
\begin{equation}
 \Phi_2 = \beta\left(D_3\partial_t^3\Phi_1 
  + D_2\partial_t^2\Phi_1 + D_1\partial_t\Phi_1 +
  D_0\Phi_1 + S_2\right),
 \label{eqn:Phi2-Phi1-dPhi1}
\end{equation}
where $D_{3,2,1,0}$ and $S_2$ are given by (\ref{eqn:def-D-S2}) with $c=1$.

Note that coefficients of the master equation (\ref{eqn:master-eq-4th})
and the relation (\ref{eqn:Phi2-Phi1-dPhi1}) were obtained just up to
the leading order in the $\tilde{\epsilon}$ expansion. It is of course 
possible to seek corrections to the leading terms, but we shall not do
so in this paper for simplicity. Note also that we had to keep up to the 
fourth order in $\tilde{\epsilon}$ in the intermediate steps to obtain
the leading master equation. The reason for this is manifest: otherwise,
we would not have been able to obtain a non-vanishing coefficient of 
$\Phi_1$ in (\ref{eqn:master-eq-4th}) since $C_0=O(\tilde{\epsilon}^4)$
while the coefficient of $\partial_t^2\Phi_1$ is unity.

\subsection{Reduction to a set of two first order equations}
\label{subsec:reduction}

Although the master equation (\ref{eqn:master-eq-4th}) obtained in the
previous subsection is formally fourth order, the fourth and third order
derivative terms are actually irrelevant. To see this is easy: they are
suppressed by the cutoff scale $M$ compared with the coefficients of the
second and first derivative terms, respectively. This means that
ignoring the fourth and third order derivative terms just drop out modes
with frequency of order $M$ or higher, which are outside the regime of
validity of the effective field theory. Properties of these modes are
dependent of properties of unknown UV completion and, thus, we not only
may but also must discard these modes. In other words, we assume the
existence of a good UV completion but will never use its
properties. Hence, we obtain the second-order master equation 
%
\begin{equation}
 \partial_t^2\Phi_1 + C_1\partial_t\Phi_1 + C_0\Phi_1 = S_1,
  \label{eqn:master-eq-2nd}
\end{equation}
where $C_{1,0}$ and $S_1$ are given by (\ref{eqn:C-def}) and
(\ref{eqn:S1-def}). For the same reason, we must discard the third and
second order derivative terms in the relation
(\ref{eqn:Phi2-Phi1-dPhi1}): 
%
\begin{equation}
 \Phi_2 = \beta\left(D_1\partial_t\Phi_1 + D_0\Phi_1 + S_2\right),
  \label{eqn:Phi2-Phi1-dPhi1-1st}
\end{equation}
where $D_{1,0}$ and $S_2$ are given by (\ref{eqn:def-D-S2}) with $c=1$.

The master equation (\ref{eqn:master-eq-2nd}) and the relation
(\ref{eqn:Phi2-Phi1-dPhi1-1st}) appear to be singular when
$M^2\Sigma+2\alpha{\bf k}^2/a^2$ vanishes. This does not mean that the
system exits the regime of validity of the low energy effective theory 
but just implies that the dynamics is not described by a single master
equation. Indeed, at the would-be singularity 
$M^2\Sigma+2\alpha{\bf k}^2/a^2=0$, the second-order master equation
is reduced to $\partial_t\Phi_1+H\Phi_1=0$. This suggest that the
second-order master equation should be decomposed into a set of two
first-order equations. To make this more explicit, let us define a new
variable $\chi$ by 
%
\begin{equation}
 \chi \equiv \frac{4\left(\partial_t\Phi_1+H\Phi_1\right)}
  {M^2\Sigma+2\alpha{\bf k}^2/a^2}.
\end{equation}
With this variable and $\Phi_1$ the master equation
(\ref{eqn:master-eq-2nd}) is reduced to the following set of two
first-order equations. 
%
\begin{eqnarray}
 \partial_t\Phi_1 + H\Phi_1 & = & \frac{1}{4}
  \left(M^2\Sigma+2\alpha\frac{{\bf k}^2}{a^2}\right)\chi, \nonumber\\
 \partial_t\chi & = & 
  \left(\frac{M^2}{M_{\rm Pl}^2}
   -\frac{2}{M^2}\frac{{\bf k}^2}{a^2}\right)\Phi_1
  + S_{\chi},
  \label{eqn:master-eq-1st}
\end{eqnarray}
where the source term $S_{\chi}$ is given by 
%
\begin{equation}
 M_{\rm Pl}^2\frac{{\bf k}^2}{a^2}S_{\chi} = 
  \frac{M^2}{2M_{\rm Pl}^2}
  \left[\left(1+3c_s^2\right)\rho\Delta+3p\Gamma\right].
\end{equation}
On the other hand, the variable $\Phi_2$ is written in terms of $\Phi_1$
and $\chi$ as
%
\begin{equation}
 \Phi_2 = \beta\left[ H\chi + 
  \left(\frac{M^2}{M_{\rm Pl}^2}-\frac{2}{M^2}
   \frac{{\bf k}^2}{a^2}\right)\Phi_1 + S_2\right],
 \label{eqn:Phi2-Phi1-chi}
\end{equation}
where
%
\begin{equation}
 M_{\rm Pl}^2\frac{{\bf k}^2}{a^2}S_2 = 
  \frac{M^2}{2M_{\rm Pl}^2}(\rho\Delta+2p\Pi).
\end{equation}
The apparent singularity at $M^2\Sigma+2\alpha{\bf k}^2/a^2=0$ does not
exist in the new equations (\ref{eqn:master-eq-1st}) and
(\ref{eqn:Phi2-Phi1-chi}). As easily seen from the structure of the set
of first order equations (\ref{eqn:master-eq-1st}), the apparent
singularity in (\ref{eqn:master-eq-2nd}) and
(\ref{eqn:Phi2-Phi1-dPhi1-1st}) was just the reflection of the fact that
the evolution of the variable $\Phi_1$ is decoupled from that of $\chi$
when $M^2\Sigma+2\alpha{\bf k}^2/a^2=0$.

\subsection{Results summarized}
\label{subsec:results-perturbation}

As shown in subsection~\ref{subsec:reconstruct-V}, in ghost condensation
it is always possible to find a form of a potential for the Higgs sector 
which realizes a given cosmological history ($H(t)$, $\rho(t)$,
$p(t)$) of the FRW background, provided that the conservation
$\partial_t\rho+3H(\rho+p)=0$ is satisfied. An important point is that
the null energy condition can be violated in the Higgs sector of gravity
without introducing ghosts or any other instabilities as far as the
violation is weak enough~\cite{Creminelli:2006xe}. Note again that the
structure of the low energy effective field theory of ghost condensation
is completely determined by the symmetry breaking pattern and does not
include any ghosts. Thus, for example we can realize the phantom
($w<-1$) cosmology without a ghost. This is an explicit counter-example
against the folklore that the phantom ($w<-1$) cosmology would require a
ghost.

In this subsection we summarize the result of this section, where we
have derived a set of two first order ordinary equations governing
cosmological perturbations of the Higgs sector of gravity. The low
energy effective theory of ghost condensation includes two dimensionless 
parameters of order unity $\alpha$ ($\equiv\alpha_1+\alpha_2>0$) and
$\beta$ ($\equiv\alpha_2$) and the cutoff scale $M$. Provided that
($\alpha$, $\beta$, $M$) are fixed and a background cosmological history
($H(t)$, $\rho(t)$, $p(t)$) (satisfying the conservation equation but
possibly including deviation from general relativity) is given, the
evolution equations for cosmological perturbations are completely
determined.

Now let us start summarizing evolution equations for cosmological
perturbation. For notation the reader should refer to the second and
the third paragraphs of this section.

The metric perturbation is decomposed into the standard, general
relativity (GR) part ($\Phi_{GR}$,$\Psi_{GR}$) and the modification part
($\Phi_{mod}$,$\Psi_{mod}$) as (\ref{eqn:GR-plus-mod}), and the
modification part is written as 
%
\begin{eqnarray}
 \Phi_{mod} & = & \Phi_1 + \Phi_2
  -\beta \frac{a^2}{{\bf k}^2}\frac{p\Pi}{M_{\rm Pl}^2}, \nonumber\\
 \Psi_{mod} & = & - \Phi_1 - 2\Phi_2
   +\beta \frac{a^2}{{\bf k}^2}\frac{p\Pi}{M_{\rm Pl}^2}.
\end{eqnarray}
The variable $\Phi_1$ is given by solving 
%
\begin{eqnarray}
 \partial_t\Phi_1 + H\Phi_1 & = & \frac{1}{4}
  \left(M^2\Sigma_0+2\alpha\frac{{\bf k}^2}{a^2}\right)\chi, \nonumber\\
 \partial_t\chi & = & 
  \left(\frac{M^2}{M_{\rm Pl}^2}
   -\frac{2}{M^2}\frac{{\bf k}^2}{a^2}\right)\Phi_1
  + S_{\chi},
  \label{eqn:master-eq-1st-again}
\end{eqnarray}
where $\chi$ is an auxiliary variable, $\Sigma_0$ is given by 
%
\begin{equation}
 \Sigma_0 = -\frac{2}{M^4}
  \left[2M_{\rm Pl}^2\partial_tH+(\rho+p)\right]
\end{equation}
and represents deviation of the background evolution from general
relativity, and the source term $S_{\chi}$ is given by 
%
\begin{equation}
 M_{\rm Pl}^2\frac{{\bf k}^2}{a^2}S_{\chi} = 
  \frac{M^2}{2M_{\rm Pl}^2}
  \left[\left(1+3c_s^2\right)\rho\Delta+3p\Gamma\right].
\end{equation}
On the other hand, the variable $\Phi_2$ is written in terms of $\Phi_1$ 
and $\chi$ as
%
\begin{equation}
 \Phi_2 = \beta\left[ H\chi + 
  \left(\frac{M^2}{M_{\rm Pl}^2}-\frac{2}{M^2}
   \frac{{\bf k}^2}{a^2}\right)\Phi_1 + S_2\right],
\end{equation}
where
%
\begin{equation}
 M_{\rm Pl}^2\frac{{\bf k}^2}{a^2}S_2 = 
  \frac{M^2}{2M_{\rm Pl}^2} (\rho\Delta+2p\Pi).
\end{equation}
Note that $\Phi_2$ vanishes if $\beta=0$.

The source terms $S_{\chi}$ and $S_2$ vanish in the $M/M_{\rm Pl}\to 0$
limit. This means that the modification part ($\Phi_{mod}$,
$\Psi_{mod}$) is not induced by matter sources in this limit. In other
words, in this limit the Higgs sector of gravity is decoupled from the
gravity and the matter sectors, and the general relativity is safely
recovered.

One must be aware that these equations are valid only if the physical
momentum $\sqrt{{\bf k}^2}/a$ is sufficiently lower than the scale of
the spontaneous Lorentz breaking $M$, which plays the role of the cutoff
scale of the low energy effective theory.

The coefficient $\Sigma_0$ includes information about deviation of the
background evolution from general relativity. In usual approach, the
deviation is considered as indication of dark energy and/or dark
matter. Instead, we here replace those dark components with the Higgs
mechanism of gravity, i.e. ghost condensation. The cases with
$\Sigma_0>0$ and $\Sigma_0<0$ correspond to dark components with 
$w>-1$ and $w<-1$, respectively. In the usual approach, a dark component
with $w<-1$ is called phantom and is thought to be associated with
ghosts, i.e. excitations with wrong-sign kinetic energy. On the other
hand, as explained in Sec.~\ref{sec:intro}, there is no ghost in the
ghost condensation. The set of equations summarized above can be applied
to both $\Sigma_0>0$ ($w>-1$) and $\Sigma_0<0$ ($w<-1$) cases, including
transition between these two regimes.

\section{Concluding remark}

In the simplest setup of the ghost condensation the shift symmetry is
exact and it behaves like a cosmological constant plus cold dark matter
for homogeneous, isotropic background evolution. With soft breaking of
the shift symmetry, a shallow potential is allowed in the Higgs sector
Lagrangian. We have investigated the classical dynamics of cosmology in
ghost condensation with softly broken shift symmetry. As we have shown
explicitly, it is always possible to find a form of the potential in the
Higgs sector Lagrangian which realizes an arbitrary cosmological history
($H(t)$, $\rho(t)$) of the visible sector of the FRW background. After
showing the reconstruction method, we have derived the evolution
equation for cosmological perturbations in the Higgs phase of gravity.

The strongest evidence for the accelerating expansion of the universe
today comes from the supernova distance-redshift relation. Using this
kind of geometrical information of the universe, it is in principle
possible to reconstruct the potential for the Higgs sector by the method
developed in subsection~\ref{subsec:reconstruct-V} of this paper. Once
this is done, the theory acquires predictive power. By using the
formalism of the cosmological perturbations summarized in 
subsection~\ref{subsec:results-perturbation}, this theory can be tested
by dynamical information of large scale structure in the universe such
as cosmic microwave background anisotropy, weak gravitational lensing
and galaxy clustering.

Results in this paper have been obtained within the regime of validity
of the effective field theory, whose structure can be determined solely
by the symmetry breaking pattern. Therefore, while we have used a
covariant $4D$ action of a scalar field as a tool for calculation, the
final results summarized in subsection~\ref{subsec:results-perturbation}
should be universal and independent of the way the Higgs phase of
gravity is realized. The same equations should hold as far as the
symmetry breaking pattern is the same, with or without a scalar field.

\section*{Acknowledgements}

The author would like to thank B.~Feng, T.~Hiramatsu, M.~A.~Luty,
S.~Saito, A.~Shirata, Y.~Suto, A.~Taruya, K.~Yahata and J.~Yokoyama for
useful discussions and/or comments. This work was in part supported by
MEXT through a Grant-in-Aid for Young Scientists (B) No.~17740134.

\appendix{Appendix}

\subsection{Simple example: Minkowski and de Sitter backgrounds}
\label{app:example}

In this appendix, as a simple application of the formula summarized in
subsection~\ref{subsec:results-perturbation}, we consider modification
of gravity in Minkowski and de Sitter backgrounds. This is the situation
considered in ref.~\cite{Arkani-Hamed:2003uy}. 

For simplicity we set $\beta=0$. In this case $\Phi_2=0$ and the
modification part of the metric perturbation is 
%
\begin{equation}
 \Phi_{mod} = -\Psi_{mod} = \Phi_1. 
\end{equation}

By setting 
%
\begin{equation}
 \Sigma_0 = 0, \quad H=H_0 \ (= \mbox{const.})
\end{equation}
and 
%
\begin{equation}
 \rho\Delta = \delta\rho, \quad
  c_s^2=\Gamma=\Pi = 0,
\end{equation}
the set of first order equations (\ref{eqn:master-eq-1st-again}) is
reduced to 
%
\begin{eqnarray}
 \partial_t\Phi_{mod} +H_0\Phi_{mod} & = & 
  \frac{\alpha}{2}\frac{{\bf k}^2}{a^2}\chi, \nonumber\\
 \partial_t\chi & = & 
  \left(\frac{M^2}{M_{\rm Pl}^2}
   -\frac{2}{M^2}\frac{{\bf k}^2}{a^2}\right)\Phi_{mod} + S_{\chi}, 
\end{eqnarray}
where the source term is now given by 
%
\begin{equation}
 S_{\chi} = 
  \frac{M^2}{2M_{\rm Pl}^4}\frac{a^2}{{\bf k}^2}\delta\rho
  =  \frac{M^2}{M_{\rm Pl}^2}\Phi_{GR}.
\end{equation}
By eliminating $\chi$ from these equations, we obtain
%
\begin{equation}
 \partial_t^2\Phi_{mod} +3H_0\partial_t\Phi_{mod}
  + \left(\frac{\alpha}{M^2}\frac{{\bf k}^4}{a^4}
     - \frac{\alpha M^2}{2M_{\rm Pl}^2}\frac{{\bf k}^2}{a^2}
     + 2H_0^2 \right)\Phi_{mod}
  = \frac{\alpha M^2}{2M_{\rm Pl}^2}\frac{{\bf k}^2}{a^2}\Phi_{GR}.
\end{equation}
This is eq.~(8.27) of ref.~\cite{Arkani-Hamed:2003uy}~\footnote{
Note that $\alpha^2$ in \cite{Arkani-Hamed:2003uy} corresponds to
$\alpha$ in the present paper.} 
and explicitly shows that in the $M/M_{\rm Pl}\to 0$ limit, the GR part
$\Phi_{GR}$ ceases to act as the source of the modification part
$\Phi_{mod}$ and the general relativity is safely recovered. By
introducing the length and time scales $r_c$ and $t_c$ as  
%
\begin{equation}
 r_c = \frac{\sqrt{2}M_{\rm Pl}}{M^2}, \quad
  t_c = \frac{2M_{\rm Pl}^2}{\sqrt{\alpha} M^3}, 
\end{equation}
This equation is rewritten as
%
\begin{equation}
 \partial_t^2\Phi_{mod} +3H_0\partial_t\Phi_{mod}
  + \left(\frac{r_c^4}{t_c^2}\frac{{\bf k}^4}{a^4}
     - \frac{r_c^2}{t_c^2}\frac{{\bf k}^2}{a^2}
     + 2H_0^2\right)\Phi_{mod}
  = \frac{r_c^2}{t_c^2}\frac{{\bf k}^2}{a^2}\Phi_{GR}.
  \label{eqn:de-Sitter-result}
\end{equation}

In the Minkowski spacetime, by setting $H_0=0$ and $a=1$ in
(\ref{eqn:de-Sitter-result}), it is easily seen that modes with the
length scale $\sim r_c$ are unstable and that the time scale of the 
instability is $\sim t_c$. This is the analogue of the Jeans instability
found in ref.~\cite{Arkani-Hamed:2003uy} and explained in
Sec.~\ref{sec:intro}. What is different from the usual Jeans instability
is that this behavior in the linearized level persists even in Minkowski
spacetime~\footnote{
The dispersion relation for a usual fluid with the background energy
density $\rho_0$ is $\omega^2=c_s^2{\bf k}^2-\omega_J^2$, where $c_s$ is
the sound speed and $\omega_J^2=4\pi G_N\rho_0$. Long-scale modes with
$c_s^2{\bf k}^2<\omega_J^2$ have instability (Jeans instability) and
contribute to the structure formation. In the Minkowski background,
where $\rho_0=0$, $\omega_J^2$ vanishes and the Jeans instability for
the usual fluid disappears.} 
and modifies the linearized gravitational potential.

It is also easy to see from (\ref{eqn:de-Sitter-result}) that the Jeans
instability disappears when the Hubble expansion rate $H_0$ is larger
than a critical value $H_c\sim 1/t_c$. Thus, the onset of the IR
modification starts at the time when the Hubble expansion rate becomes
as low as $H_c$.


\end{document}